\documentclass[prl, twocolumn, superscriptaddress]{revtex4}

\usepackage{amssymb}
\usepackage{amsmath}
\usepackage{epsfig}
\usepackage{color}

\newcommand{\kB}{k_{\mathrm{B}}}
\newcommand{\kT}{\kB T}
\newcommand{\vek}[1]{\boldsymbol{#1}}   
\newcommand{\dif}{\mathrm{d}}           
\newcommand{\abs}[1]{\left|#1\right|}

\newcommand{\eeeexp}[1]{\exp \left( #1 \right)}
\newcommand{\mean}[1]{\left<#1\right>}
\newcommand{\gtsim}{\protect\raisebox{-0.5ex}{$\:\stackrel{\textstyle >}
        {\sim}\:$}}

\newcommand{\hhat}{\hat{h}}

\begin{document}
\title{Large-scale simulations of fluctuating biological membranes}

\author{Andrea Pasqua\footnote{\label{equally}These authors contributed
    equally to this work.}}
\affiliation{Department of Chemistry, University of California, Berkeley, CA
  94720}

\author{Lutz Maibaum\textsuperscript{\ref{equally}}}
\affiliation{Department of Chemistry, University of California, Berkeley, CA
  94720}
\affiliation{Chemical Sciences Division, Lawrence Berkeley National
  Laboratory, Berkeley, CA 94720}

\author {George Oster}
\affiliation{Department of Molecular and Cellular Biology, University of California, Berkeley, CA
  94720}

\author{Daniel A. Fletcher}
\affiliation{Department of Bioengineering, University of California, Berkeley, CA
  94720}
\affiliation{Physical Biosciences Division, Lawrence Berkeley National
  Laboratory, Berkeley, CA 94720}

\author{Phillip L. Geissler}
\affiliation{Department of Chemistry, University of California, Berkeley, CA
  94720}
\affiliation{Chemical Sciences Division, Lawrence Berkeley National
  Laboratory, Berkeley, CA 94720}
\affiliation{Physical Biosciences Division, Lawrence Berkeley National
  Laboratory, Berkeley, CA 94720}

\begin{abstract}
  We present a simple, and physically motivated, coarse-grained model of
  a lipid bilayer, suited for micron scale computer simulations. Each
  $\approx 25 \mathrm{nm}^2$ patch of bilayer is represented by a
  spherical particle. Mimicking forces of hydrophobic association,
  multi-particle interactions suppress the exposure of each sphere's
  equator to its implicit solvent surroundings. The requirement of high
  equatorial density stabilizes two-dimensional structures without
  necessitating crystalline order, allowing us to match both the
  elasticity and fluidity of natural lipid membranes. We illustrate the
  model's versatility and realism by characterizing a membrane's response to
  a prodding nanorod.
\end{abstract}
\maketitle

\section{Introduction}

Lipid bilayers form the basis of biological membranes. Integral
membrane proteins and a variety of small molecules are embedded in
this two-dimensional fluid, which is stabilized by hydrophobic
interactions: the bilayer structure effectively shields the lipid's
hydrophobic alkane chains from exposure to the aqueous
solvent. Despite this complexity on the molecular level, biological
membranes at large length scales are well characterized by
surprisingly few material properties. In particular, their behavior on
large length scales is consistent with that of crude elastic models,
which take as input only a bending rigidity, i.e., the membrane's
resistance to smooth shape deformations. It is unclear below what
scale such representations become inappropriate, in part due to
computational difficulties of incorporating thermal fluctuations,
which invariably gain importance as the scale of observation is
reduced from the macroscopic. At the opposite extreme computer
simulations of lipid bilayers at atomistic resolution provide detailed
access to the spectrum of thermal fluctuations. They are limited,
however, to length and time scales of tens of nanometers and hundreds of
nanoseconds,  respectively~\cite{Brannigan06}.

Many important biological phenomena involve membrane deformations over
several micrometers and span seconds or even minutes. They thus occur
on length scales intermediate between those natural to elastic
continuum models and to atomistic representations. In attempts to
bridge this gap, a large variety of simplified models for 
interactions between lipid molecules have been proposed in the literature. These range from
systematically coarse-grained systems~\cite{Shelley01, Marrink04} to
solvent--free heuristic models~\cite{Cooke05a, Brannigan05}; a recent review can be
found in Ref.~\onlinecite{Venturoli06}. Common to these models is an
attempt to mimic the amphipathic character of individual lipid
molecules.  While considerably less expensive than atomistic
simulations in their computational demands, these approaches are still
limited in the scope of fluctuations and response they can feasibly
capture.

Extending computer simulations to examine large scale behaviors
such as aggregation of membrane-associated proteins or
deformations induced by growth of an actin network
would appear to require coarse-graining beyond the scale of individual lipid
molecules. Several such approaches have been proposed~\cite{Gompper00, Noguchi06a, Ayton08}.
Most represent in a discrete way the fluctuations implied by Helfrich's continuum model of
an elastic sheet~\cite{Helfrich73, Safran}. These require estimating
local curvature of a discretized manifold, which can be both numerically unstable
and taxing to implement.  Furthermore, these approaches do not attempt to make
a close connection with the physical forces responsible for membrane cohesion
and elasticity.

In this work we propose a new simulation model at the many-lipid
scale that follows transparently from the statistical mechanics of
hydrophobicity and the underlying membrane thermodynamics. 
Despite its simplicity, our model successfully captures
several important properties of lipid bilayer physics, such as
intrinsic fluidity and the ability to spontaneously assemble into
two--dimensional sheets.  It is furthermore sufficiently versatile to
reproduce a wide range of biologically relevant elastic properties.

In our model, we envision the lipid bilayer as a collection of small membrane
patches of size $d \approx 5 \, \mathrm{nm}$, roughly the thickness of a typical bilayer~\cite{MolecularBiologyOfTheCell}, each
comprising $\sim 100$ phopholipid molecules.  For geometric
simplicity, we represent each patch as a volume-excluding sphere with
an axis of rotational symmetry pointing from one polar head group
region to the other.  Cohesion of such patches is due of course to the
presence of water: Exposing the hydrophobic portion (i.e., the
equatorial region of our model spheres) to solvent incurs a free
energetic cost, while exposing the hydrophilic portion (i.e., the
polar caps of our model spheres) is thermodynamically advantageous. At
length scales $\gtsim 1$~nm both of these contributions should be
proportional to the exposed area~\cite{Chandler05}. Remarkably, these
considerations alone are sufficient to successfully mimic the
flexibility and fluidity of natural bilayers.

The model we have developed from these simple physical notions
resembles in some respects one reported long ago by Leibler and
coworkers~\cite{Drouffe91}. They similarly considered association of spherical units
each representing a bilayer patch comparable in size to the membrane's
thickness. The anisotropic interactions acting among their particles,
however, were devised not to reflect pertinent thermodynamic driving
forces at this length scale, but instead to foster formation of fluid
elastic sheets. In our view the potential energy function used in that
work would be difficult to motivate on microscopic grounds, and no
attempt was made in Ref.~\onlinecite{Drouffe91} to do so. For this reason we expect that
our approach will generalize more naturally to describe scenarios that
involve membrane properties beyond long-wavelength fluidity and
elasticity.

As a more immediate and practical justification for our new approach,
we found that the original implementation of the model, as described in Ref.~\onlinecite{Drouffe91}, does not
yield stable two--dimensional structures. In the Appendix we
detail a revised version of that model which is consistent with
previously reported properties. But even in this case the model
membrane's bending rigidity is atypically small for biophysical systems.

\section{Model}

For a collection of
$N$ particles, each representing a patch of lipid bilayer, we adopt the energy function
\begin{equation}\label{U}
  \mathcal{U} = \mathcal{U}_{\mathrm{HC}} + \epsilon \sum_{i=1}^N A_{\mathrm{eq}}(n^{(i)}_\mathrm{eq})
  - A_{\mathrm{pol}}(n_{\mathrm{pol}}^{(i)}),
\end{equation}
where the hard--core potential
$\mathcal{U}_{\mathrm{HC}}$ enforces the constraint that the separation
between any two particles is at least $d$.
The quantities $n^{(i)}_\mathrm{eq}$ and $n^{(i)}_\mathrm{pol}$ characterize,
respectively, the equatorial and polar coordination numbers of
particle $i$. The functions $A_{\mathrm{eq}}(n)$ and
$A_{\mathrm{pol}}(n)$ determine solvent exposure of these two regions
based on their coordination. The positive constant $\epsilon$ sets the
scale of these solvent-mediated interactions. 
Based on the surface
tension between water and oil, $\gamma \approx 50 {\rm mJ/m}^2$, and
the hydrophobic surface area of a membrane patch, $A \approx 60 {\rm
  nm}^2$, we expect $\gamma A \approx 740 k_{\rm B}T$ 
to be an appropriate value for $\epsilon$.

In detail, we define the fluctuating density of a particle's
equatorial and polar neighbors as
\begin{eqnarray}
  n_\mathrm{eq}^{(i)} & = & \sum_{j\neq i}G_\mathrm{eq}(r_{ij})H_\mathrm{eq}(z_{ij}^2), \label{eq:neq}\\
  n_{\mathrm{pol}}^{(i)} & = & \sum_{j\neq i}G_{\mathrm{pol}}(r_{ij})H_{\mathrm{pol}}(z_{ij}^2).\label{eq:npol}
\end{eqnarray}
The contributions of particle $j$ to the coordination densities of
particle $i$ are thus determined both by the distance $r_{ij}=|{\bf
  r}_{ij}|$ between their centers, and by the normalized projection
$z_{ij} = \mathbf{r}_{ij}\cdot\mathbf{\hat{d}}_i/r_{ij}$ of their
separation vector $\mathbf{r}_{ij}$ onto the axis of particle $i$
(which points along the unit vector $\mathbf{\hat{d}}_i$).  They are
attenuated by the functions
\begin{equation}
  G_\mathrm{eq}(r)=G_\mathrm{pol}(r)=\left\{
    \begin{array}{ll}
      1, & \mathrm{if~} r\le r_a ,\\
      \frac{r^2_b-r^2}{r^2_b-r^2_a}, & \mathrm{if~} r_a < r \le r_b , \\
      0, & \mathrm{otherwise},
    \end{array}
  \right.
\end{equation}
and
\begin{equation}
  H_\mathrm{eq}(z^2)=1-H_\mathrm{pol}(z^2)=\left\{
    \begin{array}{ll}
      1, & \mathrm{if~} z^2 \le z^2_a ,\\
      \frac{z^2_b-z^2}{z^2_b-z^2_a}, & \mathrm{if~} z^2_a < z^2 \le z^2_b , \\
      0, & \mathrm{otherwise} ,
    \end{array}
  \right.
\end{equation}
over scales determined by parameters $r_a$, $r_b$, $z_a$, and $z_b$.
In this way, particle $j$ counts toward the equatorial (polar)
coverage of particle $i$ only if it lies near its center and not too
far off the equatorial plane (polar axis).

For a partially occluded object, such as our idealized membrane
particles when surrounded by neighbors, calculating the area
accessible to solvent molecules of finite size is a nontrivial
operation. Furthermore, this task is not sensible to carry out in
detail for the reduced representation we have chosen. Essential
features of the functions $A_{\rm eq}(n)$ and $A_{\rm pol}(n)$ are
nonetheless straightforward to ascertain: As $n$ increases, exposed
area at first declines steadily. At some value $\overline{n}$ lower
than the maximum $\overline{n}^{\rm (max)}$ permitted by steric
constraints, it should nearly vanish, since perfect close-packing is
not needed to thoroughly exclude solvent from the bilayer's interior
(or from the polar exterior).  For $n>\overline{n}$, variation in
exposed area should be very weak.  We caricature this dependence with
computationally inexpensive, piecewise linear functions:
\begin{equation}
  A_\mathrm{eq}(n)=1-\frac{n}{\bar{n}_\mathrm{eq}}\,(n<\bar{n}_\mathrm{eq}),\,\,\,0\,\,(n\ge\bar{n}_\mathrm{eq}),
\end{equation}
\begin{equation}
  A_\mathrm{pol}(n)=1-\frac{n}{\bar{n}_\mathrm{pol}}\,(n<\bar{n}_\mathrm{pol}),\,\,\,0\,\,(n\ge\bar{n}_\mathrm{pol}).
\end{equation}
Appropriate values of $\bar{n}_\mathrm{eq}$ and $\bar{n}_\mathrm{pol}$
will depend on the relative thicknesses of hydrophobic and polar
regions, specified in our model by $z_a$ and $z_b$.

The membrane free energy we have described is markedly multi-body in
character, but in a way that is both physically meaningful and simple
to understand. Its parameters correspond intuitively to the geometry
and chemistry of constituent molecules. Below we present results of
Monte Carlo computer simulations for this model, which employed single
particle translations and rotations as trial moves, 
as well as shape deformations of the periodically replicated simulation box when
an external tension was imposed. 
For these calculations we selected
$r_a / d = 1.3, r_b / d = 1.7, z^2_a=0.05, z^2_b=0.2, \bar{n}_\mathrm{eq}=5$ and $\bar{n}_\mathrm{p}=1$,
which yield a rigidity typical of biological membranes.  By varying
these values, it is possible to tune the elasticity of the assembled
bilayer, as well as more subtle properties like internal viscosity
or rupture tension. 

\begin{figure}[tb]
\resizebox{\columnwidth}{!}{
  \includegraphics{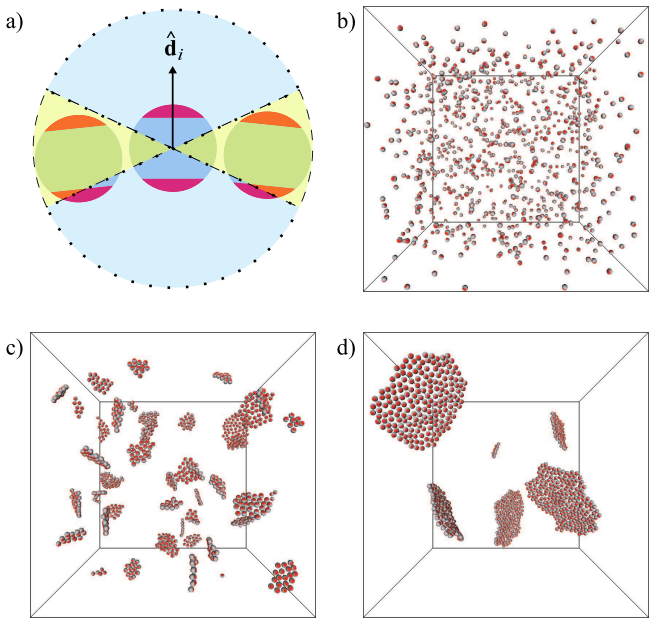}
}
  \caption{\label{fig:reform}
    (a) Illustration of our model. Each particle corresponds to a
    fragment of lipid bilayer, comprised of a central hydrophobic core 
    and two surrounding hydrophilic layers. The unit vector
    $\mathbf{\hat{d}}_i$ specifies the orientation of particle $i$. 
    Also shown are the spatial regions used to compute the numbers of
    equatorial (enclosed by dashed line) and
    polar (enclosed by dotted line) neighbors of particle $i$.
    (b) Random initial configuration for a trajectory of $N=864$
    particles. As time progresses, particles quickly form two--dimensional
    patches that continue to coarsen. Shown are snapshots after 100 (c) and
    1000 (d) Monte Carlo sweeps.
  }
\end{figure}

\section{Results}

We first demonstrate that a sheet--like configuration is indeed the
equilibrium state of our model at finite
temperature. Fig.~\ref{fig:reform} shows snapshots from a Monte Carlo
trajectory in which an initially disperse collection of membrane
particles spontaneously organizes into two--dimensional structures
that then diffuse and coalesce~\cite{footnote1}.
This coarsening process is expected to proceed
until only a single membrane sheet remains in the simulation box.
A free boundary in the resulting
structure can be avoided either by forming a membrane sheet that spans
the periodically replicated simulation box, or by adopting a
boundary--free geometry such as a spherical vesicle.

We next provide evidence that the elastic properties of our model match
quantitatively those of natural lipid bilayers. Specifically, on
length scales well beyond a particle radius $d$ (corresponding to the
bilayer's thickness), its shape fluctuations are well described by the
Helfrich model of incompressible fluid elastic
sheets~\cite{Helfrich73, Safran}, with macroscopic material properties
in accord with experimental measurements.  In Helfrich's model, a
nearly flat segment of membrane exhibits a fluctuation spectrum
\begin{equation}
  \mean{\abs{\hhat_{\vek{q}}}^2}
  =
  \frac{\kT {\cal A}}{\sigma q^2 + \kappa q^4} .
  \label{eq:hq2}
\end{equation}
Here, $\hhat_{\vek{q}} = \int_A h(\vek{x}) \eeeexp{-i
  \vek{q} \cdot \vek{x}}\,\dif\vek{x}$ is the Fourier transform of the membrane
height $h(\vek{x})$, $\vek{q}$ is a two-dimensional wavevector
conjugate to the Cartesian position $\vek{x}$ on a flat reference
membrane with area ${\cal A}$, and $\mean{.}$ denotes the equilibrium average.  For lipid mixtures common in cell
membranes, the bending rigidity $\kappa$ ranges from $10$ to $30 \,
\kT$~\cite{Rawicz00}. To the extent that the membrane's area per
molecule is constant, the surface tension $\sigma$ plays the role of
minus chemical potential and increases roughly linearly with applied
lateral tension $\tau$.

\begin{figure}[tb]
  \resizebox{\columnwidth}{!}{\includegraphics{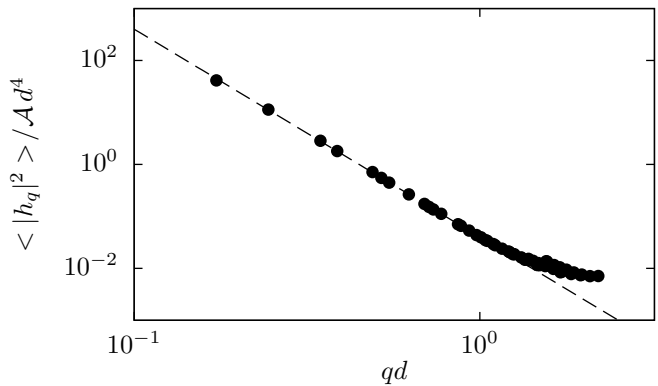}}
  \caption{\label{fig:fluctuations}
    Spectrum of height fluctuations around a flat reference state, equilibrated at
    zero lateral tension. The dashed line is a fit to the expected behavior
    \eqref{eq:hq2} with $\sigma = 0$, yielding a bending rigidity $\kappa = 25.2
    \kT$.
  }
\end{figure}

For the purpose of estimating the normal mode fluctuations of
Eq.~\ref{eq:hq2} from simulation, we wish to avoid imposing lateral
tension (requiring $\tau=0$). As a practical matter, however, it is
convenient in this calculation to prescribe a fixed box geometry (and
thus a fixed set of wavevectors). We began by performing simulations
in which the box size was allowed to fluctuate at zero lateral
tension. The resulting average box size was then adopted as a
constraint for a second set of simulations, in which we computed the
distribution of height fluctuations~\cite{footnote2}.
At large length
scales we indeed observe the characteristic
$\mean{\abs{\hhat_{\vek{q}}}^2}\propto q^{-4}$ behavior predicted by
the Helfrich model, as shown in Fig.~\ref{fig:fluctuations}. The
computed proportionality coefficient indicates a bending rigidity of
$\kappa \approx 25 \kT$, well within the range of measured
values~\cite{Rawicz00}. By manipulating the parameters $z^2_a$ and $z^2_b$,
we are able to reproduce the elastic behavior of membrane sheets over
a wide range of bending rigidities (Table \ref{tab:bendingrigidities}).

\begin{table}[tb]
  \setlength{\tabcolsep}{0.3cm}
  \begin{tabular}{r@{.}l|r@{.}l|r@{$\pm$}l}
    \multicolumn{2}{c|}{$z^2_a$} & \multicolumn{2}{c|}{$z^2_b$} & \multicolumn{2}{c}{$\kappa / \kT$} \\
    \hline
    0&01 & 0&2 & 126.8 & 0.8 \\
    0&02 & 0&2 & 63.4 & 0.5 \\
    0&05 & 0&2 & 25.2 & 0.1 \\
    0&1 & 0&2 & 12.1 & 0.1 \\
    0&4 & 0&6 & 2.1 & 0.03
  \end{tabular}
  \caption{\label{tab:bendingrigidities}Computed bending rigidities
  for different values of the model parameters $z^2_a$ and $z^2_b$.
  }
\end{table}

The advantage of our model lies in a facile ability to address
mesoscale response without sacrificing the microscopic basis of
corresponding fluctuations. As a representative biophysical example
that calls for these capabilities, we considered the resistance of a
fluctuating membrane to impingement of a nanorod oriented
perpendicular to the lipid bilayer. Experimental realizations of this
situation include extension of polymerizing actin filaments close to a
cell membrane~\cite{Liu08} and external forcing of a carbon nanotube
against a cell wall~\cite{Chen07nanoinjector}.

\begin{figure}[tb]
  \setlength{\unitlength}{\columnwidth}
  \begin{picture}(1,0.77)
    \put(0.28,0.13){\resizebox{0.69 \columnwidth}{!}{\includegraphics{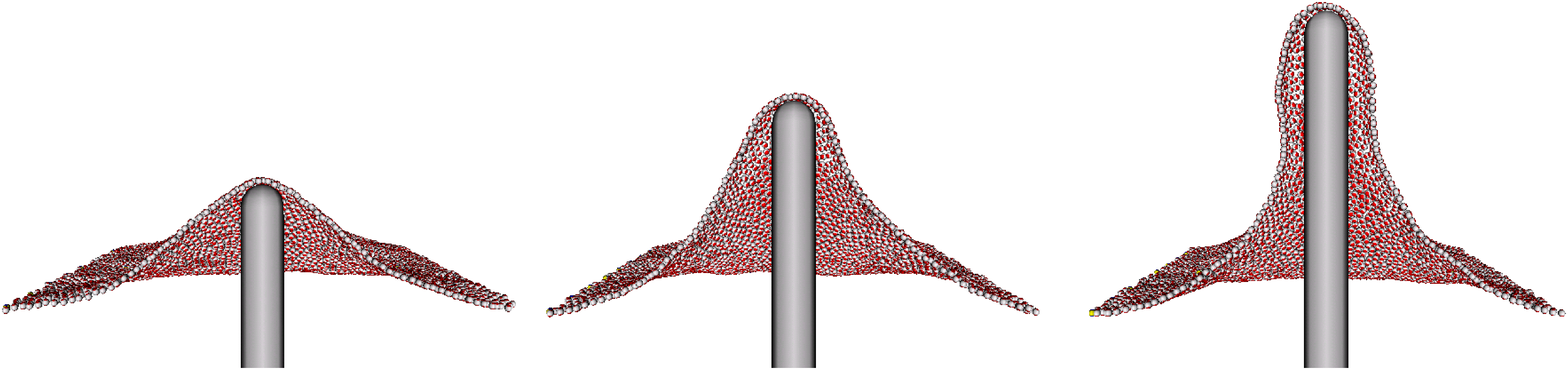}}}
    \put(0,0){\resizebox{\columnwidth}{!}{\includegraphics{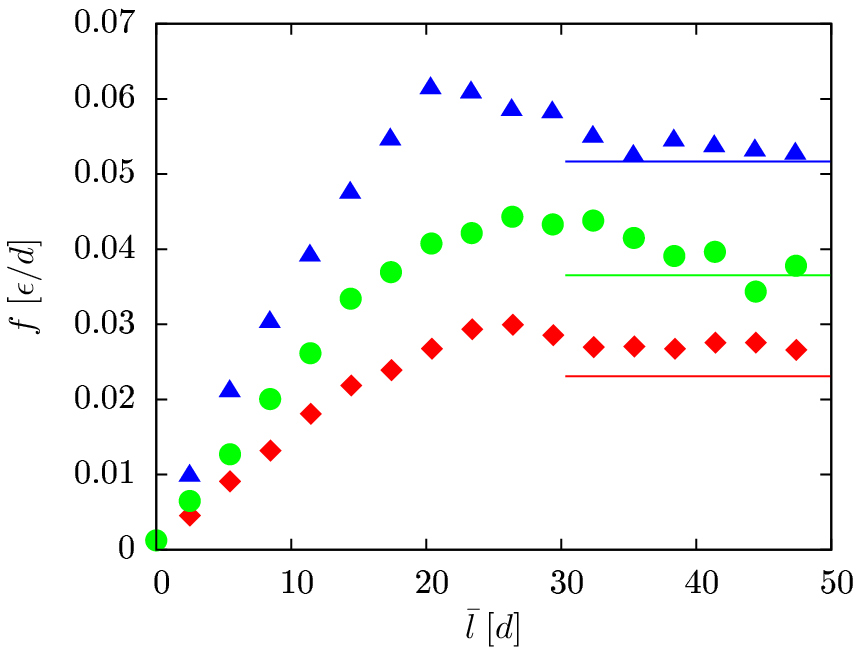}}}
  \end{picture}
  \caption{\label{fig:force}
    Restoring force as a function of protrusion length for
    three different values of the lateral tension $\tau$: 
    $0.0002 \epsilon / d^2$ (squares), 
    $0.0005 \epsilon / d^2$ (circles), and 
    $0.001 \epsilon / d^2$ (triangles). For each value of $\tau$ and 
    $l_0$ we performed computer simulations at two values of $K$ ($0.2$ and
    $1$ in units of $\epsilon / d^2$). The force was evaluated using \eqref{eq:force} at a
    length $\bar{l}$ corresponding to the arithmetic mean of the two average
    lengths $\mean{l}$ .  
    The solid lines show the result $2\pi\sqrt{2 \kappa \sigma}$ expected
    for long cylindrical protrusions in the limit of zero
    temperature with $\sigma = \tau$~\cite{Derenyi02, Atilgan06}.
    The inset shows cutaway views of typical configurations at $\tau = 0.0005 \epsilon / d^2$
    and $l_0 = 17.5, 29.5, 41.5$ (left to right), illustrating the transition
    from a global deformation to a localized, tubular protrusion. In our
    calculations the nanorod is placed at the center of the periodically
    replicated simulation box, and a small number of membrane particles are
    immobilized at the boundary to avoid overall membrane translation.
  }
\end{figure}

Insets of Fig.~\ref{fig:force} depict the reversible membrane
deformations we have studied in this context. 
The prodding nanorod is
modeled here as a volume-excluding, rigid spherocylinder of radius
$R=3d$, which defines a region in space inaccessible to the membrane
particles.  Its vertical displacement $l$ determines the size of the
membrane deformation.  We set $l=0$ for a nanorod that would contact
the membrane, in a completely flat configuration, at a single
point. To prevent global translation of the membrane when $l>0$, we
constrain the vertical positions of a small number of membrane
particles. This pinning could be viewed as a
mimicry of cytoskeletal attachments that would suppress 
overall translations in a living cell~\cite{MolecularBiologyOfTheCell}.

To calculate membrane--induced forces on the nanorod, we treat the
rod height $l$ as a dynamical variable subject to an external
potential, $V(l) = (1/2) K \left( l - l_0 \right)^2$, in addition to
the fluctuating constraints imposed by membrane particles. For several values
of $l_0$ we computed the force on a protrusion of length $\bar{l}$ from simulations as
\begin{equation}
  f(\bar{l}) = K \left[ \mean{l} - l_0 - \frac{\mean{\left(\delta l\right)^2} - \kT
      / K}{\mean{\left(\delta l\right)^2}} \left( \mean{l} - \bar{l} \right)
  \right] ,
\label{eq:force}
\end{equation}
which follows from an expansion of the membrane free energy to
quadratic order in $l$ around $\bar{l}$.

Force--extension curves for this extrusion process are plotted in
Fig.~\ref{fig:force} for three different values of $\tau$. The
restoring force initially increases in proportion to filament length,
and ultimately reaches a plateau value at large $l$. This limiting
force is in good agreement with the constant force $2\pi\sqrt{2 \kappa
  \sigma}$ predicted by the Helfrich model for the stretching
deformation of an elastic cylinder~\cite{Derenyi02, Atilgan06},
provided that $\sigma$ is dominated by lateral tension. Similar agreement is
found for the diameter of the membrane tubule (data not shown).
At intermediate extensions, $f$ exhibits a local maximum, signaling a
mechanical instability associated with the transformation between two
different classes of membrane configurations. 
This observation is consistent with recent experiments on the formation of
membrane tethers from giant vesicles~\cite{Koster05} and conclusions drawn
from continuum calculations~\cite{Derenyi02}.

\section{Outlook}

In addition to mechanical responses exemplified here by nanorod
protrusion, several other biologically relevant membrane processes could
addressed through straightforward extensions of our model. For example,
lateral demixing of multicomponent lipid bilayers could be studied by
endowing each membrane particle with a degree of freedom representing the
local composition. 
The organization of embedded proteins could be
simulated using a description of the macromolecule that is compatible with our
membrane model.
Studying time-dependent behavior would require in
addition a set of dynamical rules for updating particle
arrangements. 
Most simply, the Monte Carlo trajectories we have used
here to sample equilibrium configurations could be interpreted as an
approximation of physical time evolution. 
Integrating equations of
motion that involve derivatives of potential energy would require
smoothed versions of the interactions we have described.
Already, the mesoscale realism and computational economy of our approach
recommends its use for examining a wide range of lipid bilayer
fluctuation phenomena beyond the reach of molecular models.  

We thank Anthony Maggs for helpful discussions of the work
presented in Ref.~\onlinecite{Drouffe91}.
This work was supported by the Director, Office of Science, Office of Basic
Energy Sciences, of the U.S. Department of Energy under Contract No. DE-AC02-05CH11231.

\appendix
\section{Appendix}

The model proposed by Drouffe and coworkers in Ref.~\onlinecite{Drouffe91}
operates on the same lengthscale and focuses on the same degrees of freedom as
the model presented in this work. It employs an energy functional of the form
\begin{equation}
  \mathcal{U} = \mathcal{U}_{\mathrm{HC}} + \mathcal{U}_{\mathrm{an}} +
  \mathcal{U}_{\mathrm{den}}  .
  \label{eq:U}
\end{equation}
The first term is a pairwise repulsive energy that penalizes overlap of
different particles, which we treat as hard spheres. The second term is a pairwise anisotropic interaction
energy,
\begin{equation}
  \mathcal{U}_{\mathrm{an}}
  = \epsilon \sum_{i<j} B(r_{ij}) \left\{ 
    \eta \, (\mathbf{\hat{d}}_i \cdot \mathbf{\hat{d}}_j)^\alpha 
    + g\left(z_{ij}^2\right)
    + g\left(z_{ji}^2\right)
  \right\} ,
\label{eq:Uan}
\end{equation}
where $B(r_{ij})$ is a positive weight function that limits the range of the
interaction to $\approx 2d$, $\eta = 1$ in Ref.~\onlinecite{Drouffe91}, $\alpha \in \{1,2\}$ depending
whether the particles are symmetric, and
\begin{equation}
  g(x) = 0.75 x^3 + 0.25 x + 0.8
\end{equation}
is a monotonically increasing function over the range of its argument $x \in [0,1]$.

The last term in \eqref{eq:U} is a multibody potential that favors
configurations in which each particle is surrounded by 6 nearest neighbors,
\begin{equation}
  \mathcal{U}_{\mathrm{den}} = \epsilon \sum_i C (\rho_i),
 \label{eq:Uden}
\end{equation}
where $C(\rho_i) = (\rho_i - 6)^2 $, $\rho_i = \sum_{j \neq i} h(r_{ij})$, and
$h(r_{ij})$ is a weight function with unit value for small particle separations.

To implement this model one must specify the dependence of $B(r)$ and
$h(r)$ on inter-particle distance, which was done only graphically in
Ref.~\onlinecite{Drouffe91}. We assign these quantities simple functional forms that are
piecewise linear in $r^2$,
\begin{eqnarray}
  B(r) & =& \left\{
    \begin{array}{ll}
      1.5, & \mathrm{if~} r\le 1.85 ,\\
      \frac{1.5 (2.0^2-r^2)}{2.0^2-1.85^2}, & \mathrm{if~} 1.85<r\le 2.0 , \\
      0, & \mathrm{otherwise} ,
    \end{array}
  \right. \\
  h(r) & =& \left\{
    \begin{array}{ll}
      1, & \mathrm{if~} r\le 1.6 ,\\
      \frac{1.95^2-r^2}{1.95^2-1.6^2}, & \mathrm{if~} 1.6 < r \le 1.95 , \\
      0, & \mathrm{otherwise} .
    \end{array}
  \right. 
\label{eq:h}
\end{eqnarray}
Values of $B$ and $h$ resulting from this choice closely resemble those
plotted in Ref.~\onlinecite{Drouffe91}.

\begin{figure}[tb]
  \resizebox{\columnwidth}{!}{
    \includegraphics{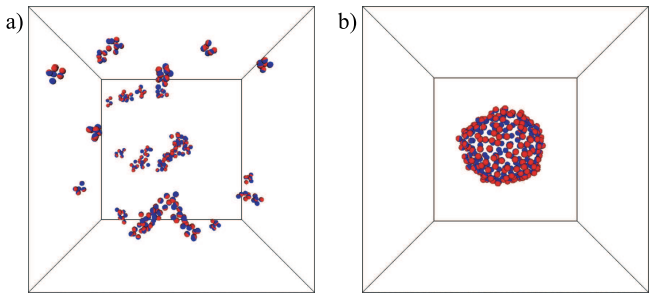}
}
  \caption{\label{fig:models}
    (a) Typical configuration of the membrane model developed by Drouffe and
    coworkers ($\alpha=\eta=1$, $T=1.6\epsilon$). A flat sheet of 256 particles with parallel orientation vectors
    was used as the initial configuration. This structure dissolves quickly
    into a disordered collection of small clusters. (b) Final
    configuration of a trajectory starting from the same initial state, but
    employing an alternative parameter set ($\alpha =1$, $\eta=-1$,
    $T=0.5\epsilon$). The initially flat sheet remains stable, and closes
    to form a vesicle.
  }
\end{figure}

We performed Monte Carlo computer simulations of the model defined by
\eqref{eq:U}--\eqref{eq:h} for $\alpha = 1$ at a temperature $T = 1.6
\epsilon$. For these conditions, Drouffe and coworkers report the spontaneous assembly of
randomly distributed and oriented particles into a two-dimensional sheet, in which the orientation vectors of neighboring particles are nearly parallel.
However, in our simulations we find that for these
parameter values the system forms many small clusters of particles without
apparent internal order. Similar configurations are obtained when a completely
ordered sheet is used as the initial condition (Fig. \ref{fig:models}a), which indicates that the inability to form stable membranes is thermodynamic in origin, and not due to dynamical artefacts caused by the Monte Carlo method.

The lack of stable membrane-like structure in this published version
of the model is straightforward both to understand and to remedy. When
$\eta=1$, the first term in the brackets of Eq.~\eqref{eq:Uan} actually disfavors
parallel alignment of neighboring particles' orientational vectors;
this energy contribution is minimum when axes of adjacent particles
are antiparallel (in the case $\alpha=1$) or perpendicular
($\alpha=2$). Suspecting a typographical error in Ref.~\onlinecite{Drouffe91}, we set
$\eta=-1$ in order to instead favor the desired alignment. Equilibrium
states resulting from this modification, however, still feature a
collection of many small aggregates of particles at the temperature
$T=1.6$ reported in Ref.~\onlinecite{Drouffe91}. In this case the adhesive interaction
 \eqref{eq:Uden} is insufficient to overcome the loss in entropy associated
 with forming a single membrane sheet. Only after reducing the temperature to
 $T = 0.5 \epsilon$ were we able to observe stable sheet-like structures and
 vesicles in our simulations (Fig. \ref{fig:models}b). This
 modified set of model parameters results in membrane sheets with a bending
 rigidity $\kappa = 4.9 \pm 0.1 \kT$, which is smaller than 
 values measured for biophysical phospholipid bilayers~\cite{Rawicz00}.

\begin{thebibliography}{10}
\bibitem{footnote1}
The elastic behavior of the model membrane is dominated by entropic effects,
yielding a nearly temperature--independent reduced bending rigidity $\kappa /
\kT$. Translation and rotation of finite membrane patches, on the
other hand, proceed by displacements of individual particles, which are
suppressed at low temperatures. To accelerate the self--assembly process 
the trajectory shown in Fig.~\ref{fig:reform} was generated at a higher
temperature $\epsilon = 25 \kT$.
\bibitem{footnote2}
To compute the
fluctuation spectrum of a membrane that spans the simulation box in
the $xy$--plane we define the height function $ h(x,y) = \sum_{i=1}^N
z_i \Theta \left( x_i - x \right) \Theta \left( y_i - y \right) /
\sum_{i=1}^N \Theta \left( x_i - x \right) \Theta \left( y_i - y
\right) $ where $\Theta(x) = 1$ if $\abs{x} \leq \Delta/2$ and zero
otherwise.  Evaluating this function for points $(x,y)$ that form a
square grid of resolution $\Delta$ and taking the two--dimensional
discrete Fourier transform, we obtain the Fourier amplitude
$\hhat_{\vek{q}}$.  This procedure generates a systematic bias at
large wavevectors~\cite{Cooke05b}. Taking this bias into account does
not change the value of $\kappa$ for our model.
\end{thebibliography}
\end{document}